\documentstyle[amstex,epsf,preprint,aps,amssymb]{revtex}

\begin{document}
\title{{\it Ab~initio} study of Curie temperatures of diluted 
III-V magnetic semiconductors }
\author{J.~Kudrnovsk\'y$^{1,4}$, I.~Turek$^{2,3,4}$, V.~Drchal$^{1,4}$,
J.~Ma\v{s}ek$^{1}$, F.~M\'aca$^{1,4}$, and P.~Weinberger$^{4}$}
\address{$^1$ Institute of Physics, Academy of Sciences of the Czech
Republic, \\
Na Slovance 2, CZ-182 21 Prague 8, Czech Republic \\
$^2$ Institute of Physics of Materials, Academy of Sciences of the Czech
Republic, \\
\v{Z}i\v{z}kova 22, CZ-616 62 Brno, Czech Republic \\
$^3$ Department of Electronic Structures, Charles University, Ke Karlovu 5,\\
CZ-121 16 Prague, Czech Republic \\
$^4$ Center for Computational Materials Science, Technical University of
Vienna, \\
Getreidemarkt 9, A-1060 Vienna, Austria}
\maketitle

\begin{abstract}
The electronic structure of diluted (Ga,Mn)As magnetic semiconductors
in the presence of As-antisites and magnetic disorder is studied within 
the framework of the local spin density approximation.
Both the chemical and magnetic disorders are treated using the
coherent potential approximation.
A ground state with partial disorder in the local moments and with
a reduced total magnetic moment appears in the presence of As-antisites.
We first estimate the Curie temperature $T_{c}$ from total energy
differences between the ferromagnetic and the paramagnetic state by
identifying these with the corresponding energy difference in a 
classical Heisenberg model.
A more systematic approach within the framework of the mean-field 
approximation to estimate $T_{c}$ consists in an evaluation of the 
effective exchange fields acting on the magnetic moment at a given site.
The presence of As-antisites strongly reduces the Curie temperature.
The results indicate that the effect of impurities on the electronic 
structure cannot be neglected and influences the Curie temperature
non-negligibly.
A comparison of the calculated Curie temperatures to existing experimental 
data indicates an increase of the donor concentration with the increase of 
the Mn-content.

\end{abstract}

\date{today}
\draft

\newpage

\section{INTRODUCTION}
\label{sec_Intr}

Because of their ferromagnetic properties, diluted magnetic III-V  
semiconductors (DMS) represent a new class of materials with potential 
technological applications in spintronics \cite{exp}.
The physics of DMS is in particular interesting because of the yet
unclear origin of the occurring ferromagnetism and the even less 
understood role of disorder.
It is now generally accepted that the ferromagnetic coupling in
III-V DMS containing Mn impurities is mediated via valence band
holes in the host semiconductor.
Mn atoms that substitute three-valent cations in the host act as
acceptors which in turn create holes in the valence band.
For finite concentration of Mn atoms the Fermi energy is then pinned 
within the valence band.
The ferromagnetic coupling of Mn atoms is often qualitatively explained 
by the fact that because of the small size of the corresponding hole 
Fermi surface the period of the Ruderman-Kittel-Kasua-Yosida (RKKY) 
oscillations exceeds the typical distance between Mn atoms. 

Theoretical approaches to the DMS can be divided into two groups, 
namely model studies using the effective-mass approximation,
and studies based on the detailed knowledge of the electronic
structure.
The model studies mostly employ the kinetic-exchange model (KEM) in 
the connection with a continuum approximation for the distribution
of Mn atoms and other defects \cite{kem,dietl}, a model that yields
a disorder-free problem, although it was recently refined in order
to include disorder via a supercell method by means of Monte Carlo 
simulations \cite{MC,ncol1}.
A simpler model assuming that holes hop only between Mn-sites, where
they interact with the Mn moments via phenomenological exchange 
interactions, was also developed \cite{bb}.

Electronic structure studies based on the spin-density functional
theory (DFT) in the framework of the local spin-density approximation
(LSDA) represent a further step towards a more detailed, parameter-free 
description of the properties of DMS.
One possibility of describing such system is a supercell approach 
\cite{fp1,fp2,fp3}, in which large cells are used to simulate 
semiconductor crystals with magnetic atoms and other impurities.
The supercell approach has limited applicability because it leads 
to excessively large supercells for low concentrations of impurities
characteristic for DMS.
Alternatively, one can employ Green function methods in which the 
configurational averaging is performed within the coherent potential 
approximation (CPA) \cite{cpa} and the resulting averaged Green function
can be used to determine any physical quantity of interest.
The Green function approach can be implemented, e.g., in the framework of the
Korringa-Kohn-Rostoker (KKR) method \cite{fp4,fp5} or the tight-binding
linear muffin-tin orbital method (TB-LMTO) \cite{tblmto} as in the present 
case.
DMS with Curie temperatures ($T_{c}$) in the range of the room temperature 
are needed for practical applications; the study of realistic models 
for evaluation of $T_{c}$ is thus of great importance.
The evaluation of the Curie temperature on {\it ab initio} level is
still a challenging task, in particular for complex alloy systems
like the DMS.
Conventional approaches known from model studies \cite{kem,dietl} cannot 
be directly employed for evaluation of $T_{c}$.
Therefore, a different approach has to adopted for the LSDA Hamiltonian.
It is well-known \cite{moriya} that transversal fluctuations of the local 
magnetization have to be taken into account to describe temperature-dependent 
properties of magnetic itinerant systems.
The direct evaluation of the corresponding enhanced frequency-dependent 
magnetic susceptibility \cite{savr} is very complicated, in particular
for such complex random systems like DMS. 
A tractable approach is based on the adiabatic treatment of atomic moments 
in which noncollinear configurations of magnetic moments are taken
into account.
The adiabatic approximation is well justified for magnetic atoms with large 
exchange splittings, like, e.g., Mn-atoms.
This method can be implemented either in the real space 
\cite{lie,curieb,curies} or in the reciprocal space (the frozen-magnon 
approach) \cite{sw}. 
The real-space approach can be naturally applied to random system in the
framework of the CPA while in the reciprocal-space approach the supercell 
method has to be adopted \cite{sb}.

In the real-space approach employed here the total energy is mapped onto 
a classical Heisenberg model whose statistical treatment represents 
a next step.
The mapping is further simplified by using the magnetic force-theorem
\cite{lie} which allows to use the band energy of the calculated ground 
state as a good estimate also for total-energy differences of the perturbed
ground state. 
The separated treatment of statistical degrees of freedom is another
advantage of this approach as it can be done on different levels of
sophistication.
It is one of the goals of the present paper to describe evaluation of 
the Curie temperature of DMS in the framework of the above mentioned
two-step model.

DMS containing Mn atoms, such as (Ga,Mn)As mixed crystals, are 
highly compensated systems.
The experimentally observed number of holes in the valence band is 
much lower than the concentration of Mn-impurities and indicates
the presence of other lattice defects acting as donors.
The As-antisites, i.e., As atoms on the Ga-sublattice, add two 
electrons to the valence band for each defect compensating thus 
two holes. 
Interstitial Mn atoms acting as double donors have the same effect 
\cite{fp3} as the As-antisites.
Recent theoretical studies indicate that magnetic structures with partial
disorder of local magnetic moments can lower the total energy of system
as compared to the ferromagnetic state. 
The stability of the ferromagnetic state was studied in \cite{ncol1} 
within the KEM model concluding that non-collinear ferromagnetism is
common in DMS.
A new magnetic state stabilized by the As-antisites and characterized
by a partial disorder in the orientations of the local Mn-magnetic 
moments was found recently also in first-principles studies of 
Ga$_{0.96}$Mn$_{0.04}$As alloys \cite{ncol3}.
All this indicates the great importance of various defects for the 
electronic structure, magnetic properties, and the Curie temperature 
of the DMS.
A detailed understanding of such influence calls for a parameter-free approach.
A detailed study of the influence of As-antisites is another goal of the 
present paper.

\section{ELECTRONIC STRUCTURE}
\label{sec_Elstr}

In this section we present results of electronic structure
calculations, in particular the magnetic phase diagram, magnetic
moments, and densities of states for the whole range of physically
interesting concentrations of Mn-impurities and As-antisites. 

\subsection{Model and computational details}
\label{CD} 

We have determined the electronic structure of DMS in the framework
of the {\it ab initio}  all-electron tight-binding linear muffin-tin 
orbital (TB-LMTO) method within the atomic-sphere approximation 
\cite{tblmto}.
The valence basis consists of $s$-, $p$-, and $d$-orbitals, we include 
scalar-relativistic corrections but neglect the spin-orbit effects.
The substitutional disorder due to Mn-atoms and As-antisites is 
treated in the CPA.
Although the CPA neglects local environment effects and possible 
lattice relaxations it correctly reproduces concentration dependent
trends and can also treat systems with small but finite concentrations 
of defects typical for DMS.
We introduce empty spheres \cite{es} into the interstitial positions 
of the zinc-blende GaAs semiconductor for a better space filling.
The lattice constant of the pure GaAs ($a=5.653~{\rm \AA}$) was used 
in all calculations but we have verified that we can neglect a weak 
dependence of the sample volume on defect concentrations.
We used equal Wigner-Seitz radii for all atoms and empty spheres.
Charge selfconsistency is achieved within the framework of the local
spin density approximation using the Vosko-Wilk-Nusair parametrization 
for the exchange-correlation potential \cite{VWN}.
The details of the method can be found in Ref.~\onlinecite{book}.

In general, in addition to the chemical disorder, the DMS are 
characterized also by some degree of magnetic disorder.
We treat this disorder in the framework of the disordered local
moment (DLM) method \cite{ncol3,dlm1,dlm2} which is the simplest
way of including disorder in spin orientations.
The DLM can be included in the framework of the CPA: the Mn atoms 
have collinear but random positive (Mn$^{+}$) and negative (Mn$^{-}$) 
orientations.
The corresponding concentrations, $x^{+}$ and $x^{-}$, fulfill the
condition $x=x^{+}+x^{-}$, where $x$ is the total Mn-concentration.
The degree of magnetic disorder can be characterized by the order 
parameter $r=(x^{+}-x^{-})/x$ which is directly related to the
macroscopic magnetization.
Clearly, $x^{\pm}=(1 \pm r)x/2$.
The GaAs mixed crystal with prescribed concentrations of $x$ Mn- and 
$y$ As-atoms on the Ga-sublattice can thus be treated formally
as a multicomponent
(Ga$_{1-x-y}$Mn$^{+}_{(1+r)x/2}$Mn$^{-}_{(1-r)x/2}$As$_{y}$)As alloy with
a varying parameter $r$, $0 \le r \le 1$.
In the saturated ferromagnetic (FM) state, $r=1$, all magnetic
moments are aligned in the direction of a global magnetization.
The paramagnetic (PM) state, $r=0$, is characterized by a complete 
disorder of spin directions with vanishing total magnetization.
The magnetic state with $0 < r < 1$ , the partial ferromagnetic
(pFM) state, is characterized by some degree of disorder in the 
spin-orientations \cite{ncol3}.
It should be noted that the DLM method is used here only to determine the
corresponding ground magnetic state in the presence of impurities.
The Curie temperature is, however, determined from the Heisenberg Hamiltonian 
corresponding to a given magnetic ground state where fluctuations of local
magnetization are taken into account.

\subsection{Magnetic phase diagram}
\label{MPD}

\begin{figure}[tbp]
\caption{(a) The three phases in the magnetic phase diagram of
(Ga$_{1-x-y}$Mn$_{x}$As$_{y}$)As. The dashed line separates $n$-
and $p$-type samples.
(b) The differences $E_{{\rm tot}}$ $-$ $E_{{\rm ground}}$ are plotted
as a function of the order parameter $r$ for an alloy
(Ga$_{0.95-y}$Mn$^{+}_{x^{+}}$Mn$^{-}_{x^{-}}$As$_{y}$)As,
$x^{\pm}=(1 \pm r) x/2$, $x=0.05$, and for a set of concentrations
$y$ of As-antisites.
Full circles: lattice constants of GaAs, empty circles:
experimental lattice constants (for $y=0.015$ only). }
\vspace*{1cm}
\epsfxsize=20cm \centerline{\epsffile{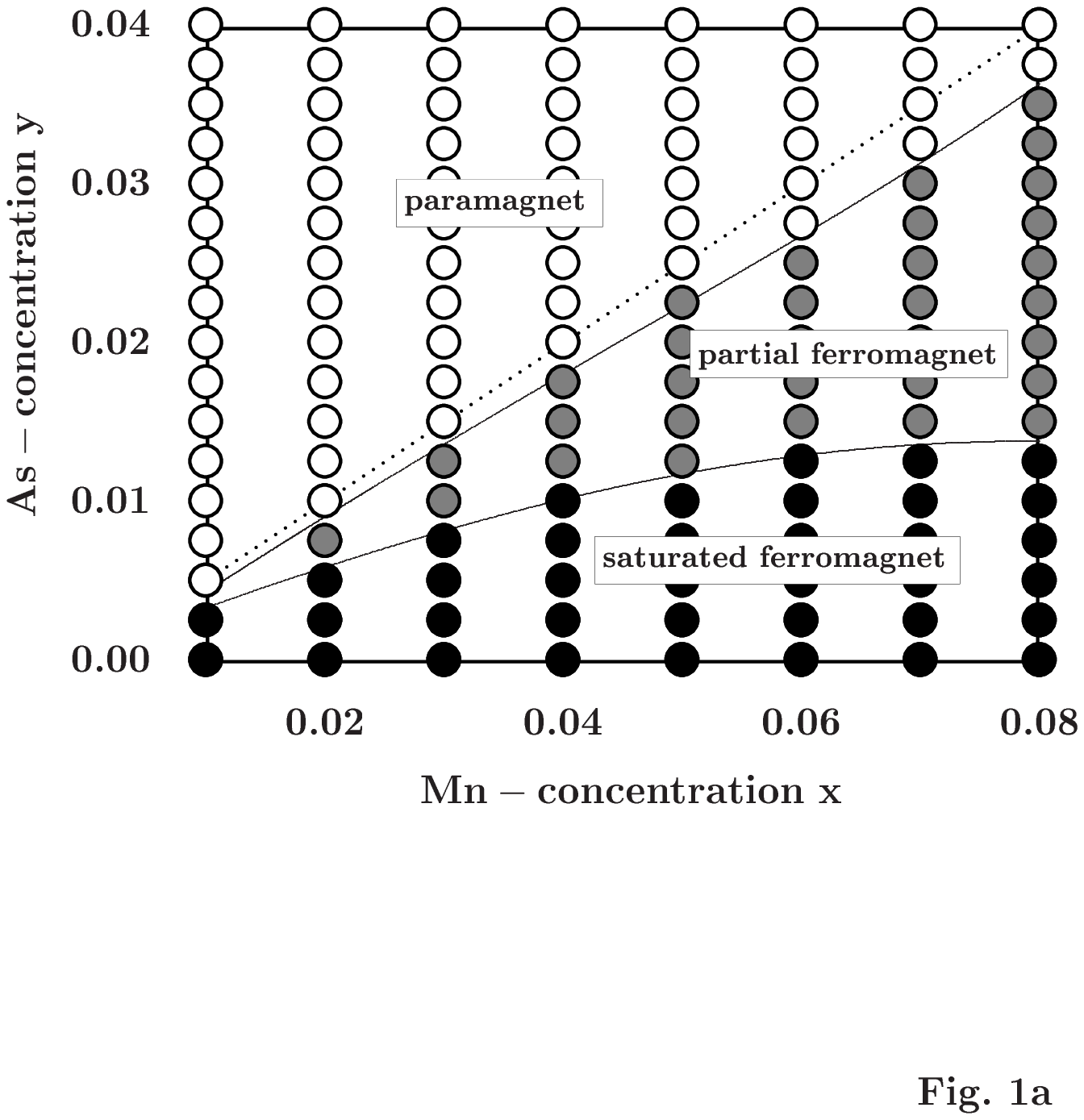}}
\newpage
\vspace*{5cm}
\epsfxsize=12cm \centerline{\epsffile{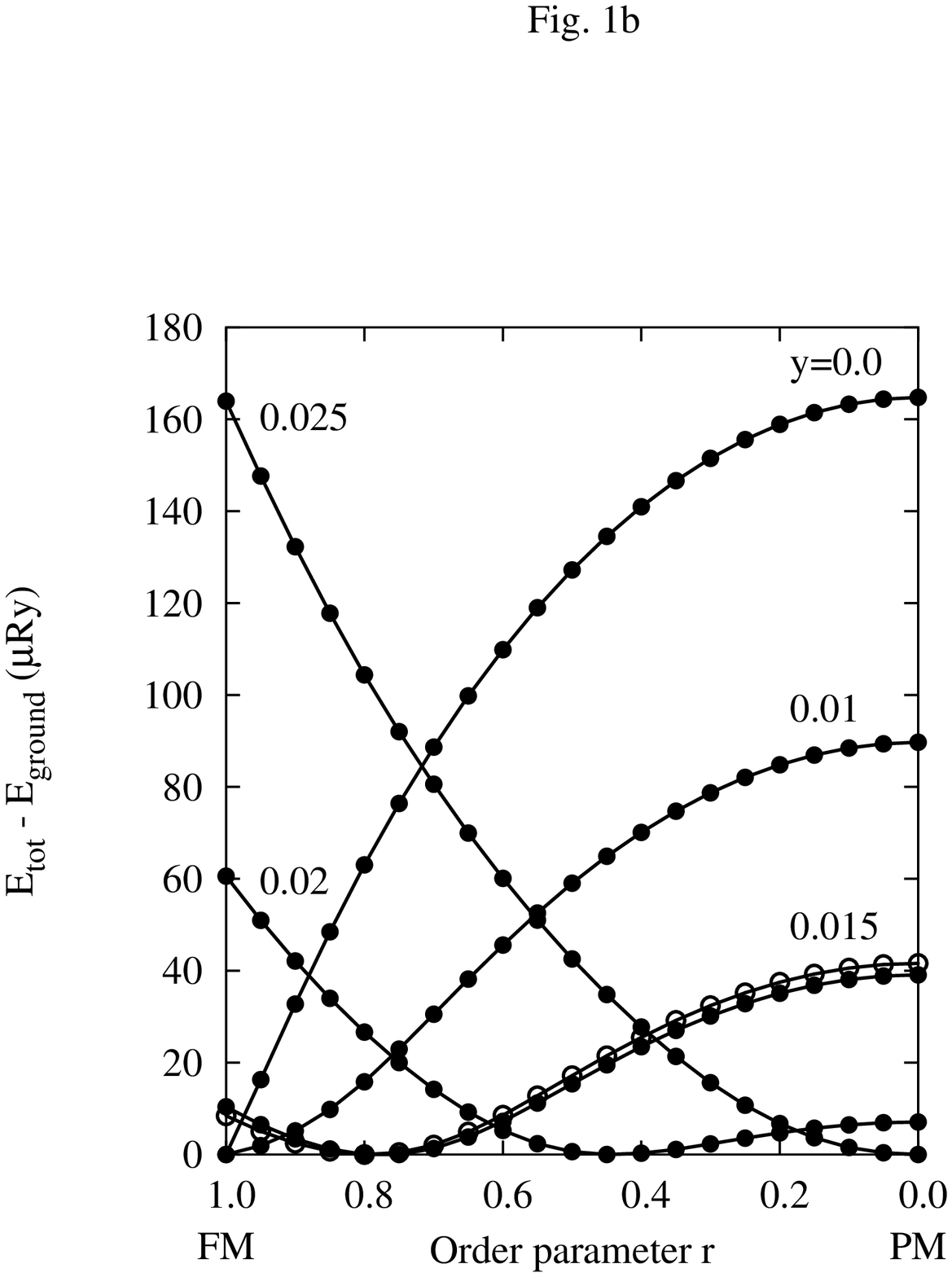}}
\label{Fig.1}
\end{figure}

We have determined the order parameter $r$ that corresponds to the ground 
state at $T=0$ for the whole range of physically relevant 
concentrations $(x,y)$ in steps of $\Delta x=0.01$ and $\Delta y=0.0025$.
In order to find the order parameter $r$ we varied for each composition 
$(x,y)$ the ratio $r$ in steps $\Delta r=0.05$ and evaluated the 
corresponding total energy.
The magnetic phase diagram determined in this way is presented in
Fig.~\ref{Fig.1}a.
It should be noted that in addition to the FM and PM states there exist 
the pFM state with a certain degree of disorder in the spin-orientations.
The pFM state, which separates the FM phase from the PM phase, appears
in the concentration region where the number of holes is reduced by 
As-antisites ($y>0$). 
Without As-antisites the ground state is the FM state. 
For $y > x/2$, i.e., in the $n$-type materials, we did not find any 
indication for an electron-induced ferromagnetism even at the highest 
studied concentrations of donors.
Further details can be read off from Fig.~\ref{Fig.1}b, in which for
a fixed Mn concentration of $x=0.05$ the total energy with respect to
the ground state is displayed as a function of the order parameter $r$.
For less than 1\% of As-antisites the ground state is a saturated
ferromagnet, for more than 2.35\% of As-antisites a paramagnet, with
a partial ferromagnet inbetween.
Empty circles for $y$=0.015 in Fig.~\ref{Fig.1}b denote results obtained
for the experimental lattice constant of (Ga$_{0.95}$Mn$_{0.05}$)As
alloy ($a=5.672~{\rm \AA}$) and demonstrates that the ground state is not
changed by a variation of the lattice constant.

\subsection{Magnetic moments}
\label{MM}

The magnetic state of disordered magnetic semiconductors can be characterized
by the local magnetic moment at the sites occupied by Mn atoms and by the
total magnetic magnetization which involves both localized moments at
magnetic impurities and the induced moments at non-magnetic atoms.
The calculations show that the total magnetic moment per unit cell is
proportional to the concentration of Mn atoms.
This is the reason why we consider the total magnetic moment normalized
to one Mn atom.

\begin{figure}[tbp]
\epsfxsize=12cm \centerline{\epsffile{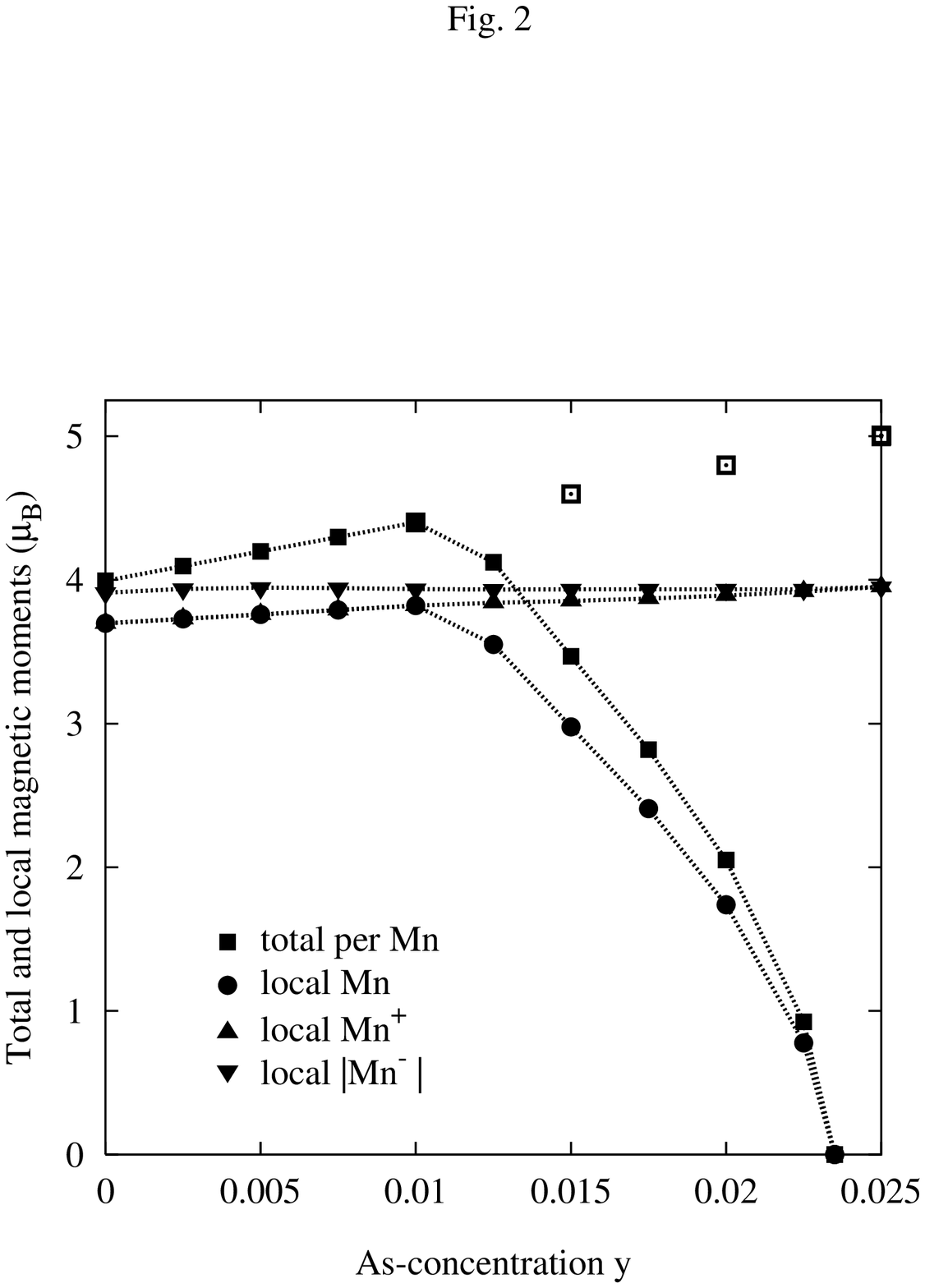}}
\caption{ Total magnetic moments and averaged local moments on Mn-atoms
for (Ga$_{0.95-y}$Mn$_{0.05}$As$_{y}$)As as a function of the concentration
of As-antisites $y$. Full symbols: the ground state, empty symbols:
the ferromagnetic state. Also plotted are the local moments on Mn$^{+}$
and Mn$^{-}$ sites in the partial ferromagnetic state
($0.011 < y < 0.0235$). In the ferromagnetic state ($y < 0.011$) the local
moments on Mn$^{+}$-sites coincide with the local Mn-moments while the local
Mn$^{-}$-moments correspond to the single-impurity limit ($x^{-}=0$
in this region).  }
\label{Fig.2}
\end{figure}

The calculated total and the local Mn-magnetic moments for 
(Ga$_{0.95-y}$Mn$_{0.05}$As$_{y}$)As are presented in
Fig.~\ref{Fig.2} as a function of the concentration $y$ of As-antisites.
The size of the local moments on Mn atoms is almost independent of the
chemical composition and depends only weakly on the direction of the
moments.
This justifies, {\it a posteriori}, the validity of the DLM model in
the present case \cite{dlm1,dlm2}.

The total magnetic moment per Mn-atom equals to 4~$\mu_B$ in the
system without As-antisites.
The integer value of the moment is a consequence of the half-metallic
character of the system.
It is in agreement with the expectation that the localized moment of
5~$\mu_B$ combines with the opposite moment of one hole in the 
majority-spin valence band.
The difference between total and local moments shows the degree 
of the magnetic polarization of the host material.
Although the induced moments on non-magnetic atoms are generally very
small, they add to a non-negligible contribution to the total magnetic
moment. 

In systems with compensating impurities, the total magnetic moment per 
Mn-atom increases linearly with $y$ and extrapolates to the value of 
5~$\mu_B$ in the completely compensated ferromagnetic state.
In reality, however, this situation is not reached because the
transition from the FM to the pFM state takes place at $y \approx 0.01$.
The transition is marked by a drop of the magnetization in the total
magnetic moment for $y > 0.01$.
The reason for the decrease of the magnetization in the pFM state is
obvious: the averaged local moment on Mn atoms, 
$m^{\rm Mn} =  x^{+}m^{{\rm Mn}^{+}}+x^{-}m^{{\rm Mn}^{-}}$, is strongly 
reduced in the presence of Mn-sites with reversed moments.
Clearly, the induced part of the magnetization also decreases 
proportionally to $m^{\rm Mn}$.

\subsection{Densities of states}
\label{DOS}

\begin{figure}[tbp]
\epsfxsize=12cm \centerline{\epsffile{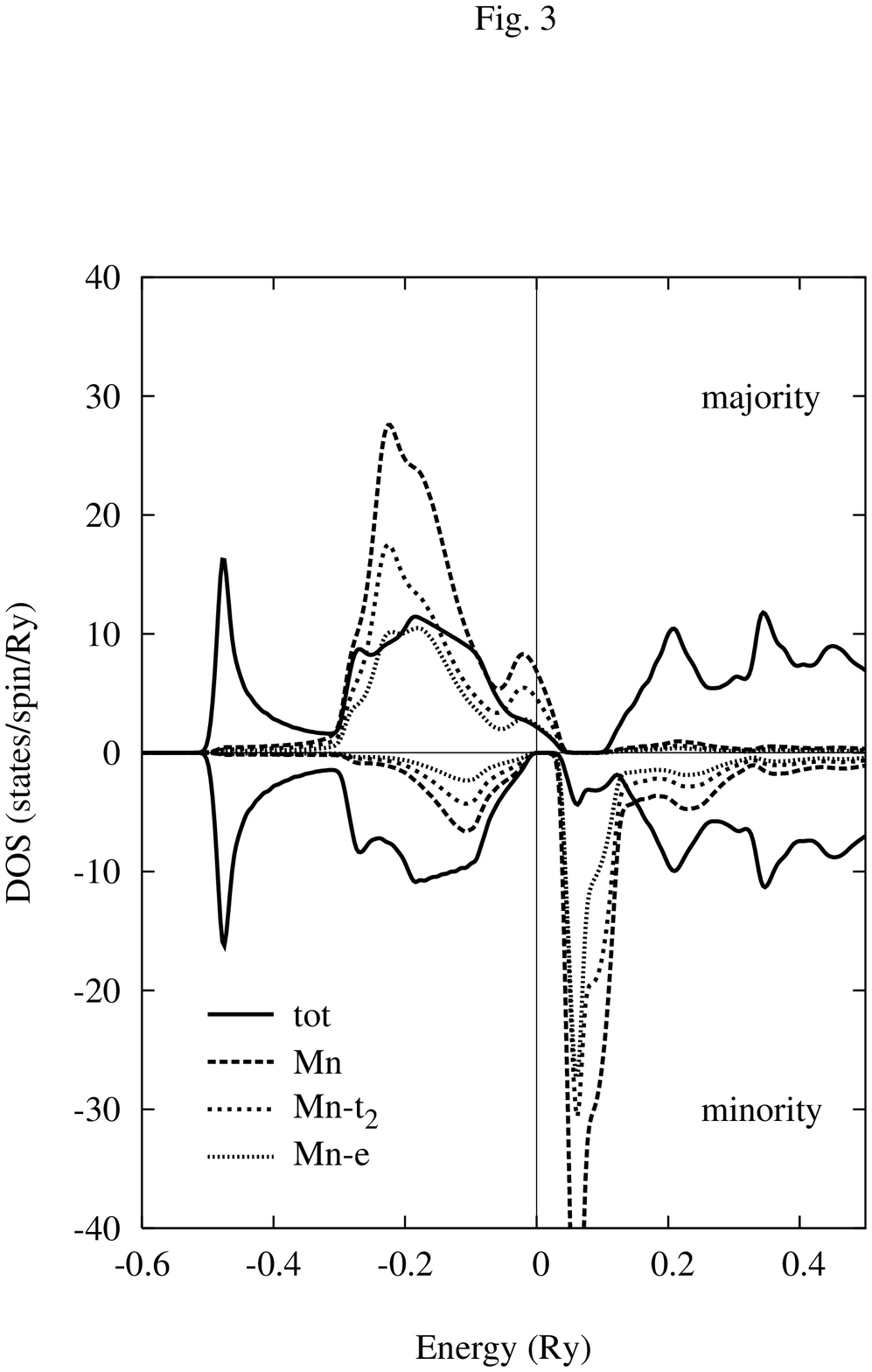}}
\caption{ Spin-dependent total densities of states and local Mn-densities
of states for the Ga-sublattice in (Ga$_{0.95}$Mn$_{0.05}$)As as
resolved with respect to the $t_{2}$- and the $e$-symmetry. The Fermi
level coincides with the energy zero. }
\label{Fig.3}
\end{figure}

\begin{figure}[tbp]
\epsfxsize=12cm \centerline{\epsffile{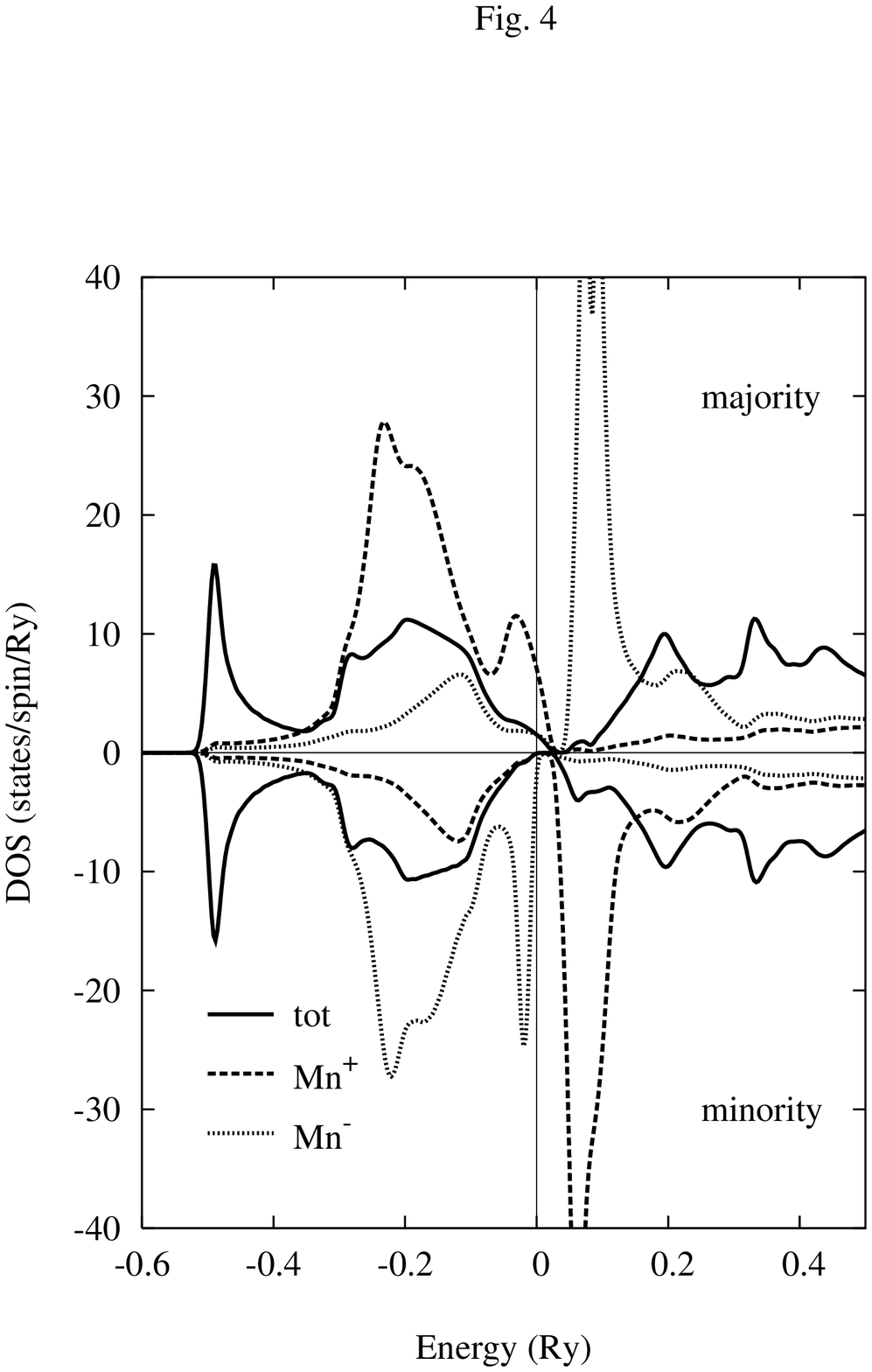}}
\caption{ Spin-dependent total densities of states and local Mn$^{+}$-
and Mn$^{-}$-densities of states for (Ga$_{0.935}$Mn$_{0.05}$As$_{0.015}$)As
in the partial ferromagnetic state. The Fermi level coincides with
the energy zero.}
\label{Fig.4}
\end{figure}

The total density of states (DOS) per unit cell and the local Mn-density of 
states for (Ga$_{0.95}$Mn$_{0.05}$)As in the FM state are presented 
in Fig.~\ref{Fig.3}.
We find a good agreement with existing results (see, e.g., 
Ref.~\onlinecite{fp5}).
In addition, we present the Mn-local DOS on the Ga-sublattice resolved 
with respect to orbital symmetries.
The different behavior of $t_{2}$- and  $e$-states which are
split due to the tetragonal environment is clearly seen.
The $t_{2}$-states just above the top of the GaAs host 
valence band were experimentally observed \cite{fp2,tckem}.
The influence of As-antisites is illustrated in Fig.~\ref{Fig.4}.
We have chosen $y=0.015$, i.e., the ground state is the pFM state
with a partial disorder in local moments (see Fig.~\ref{Fig.1}a).
The local Mn$^{+}$-DOS is quite similar to that in Fig.~\ref{Fig.3}
because $x^{+} \approx 0.044$.
On the contrary, the local Mn$^{-}$-DOS exhibits a sharp peak which is 
characteristic for a single-impurity limit ($x^{-} \approx 0.006$).
Such behavior also qualitatively explains the stability of the FM phase
for systems without As-antisites: the Fermi energy in the pFM state 
without antisites, i.e., with larger number of holes in the valence band,
would be shifted downwards and would  be situated within a sharp peak 
of the minority Mn$^{-}$ states which is energetically unfavorable.

\section{ CURIE TEMPERATURE}
\label{TC}

In this section we present two different ways of evaluating the mean-field
Curie temperature using the selfconsistent DFT calculations.
They are both implicitly based on the classical Heisenberg model,
however, without the need for an explicit evaluation of the corresponding
pair exchange interactions.

\subsection{ Simple estimate of the Curie temperature}
\label{seTC}

A simple estimate of $T_{c}$ can be obtained from DFT calculations
by relating $T_{c}$ to the total energy difference per unit cell between 
the paramagnetic and the ferromagnetic state as obtained from selfconsistent
calculations, i.e., $\Delta =E_{{\rm PM}}-E_{{\rm FM}}$.
The DFT describes only the ground state properties, therefore an additional
procedure is needed to estimate $T_{c}$.
The classical Heisenberg model 
$H=-\ \sum_{i\neq j}J_{ij}\ {\bf e}_{i}\cdot {\bf e}_{j}$, 
where ${\bf e}_{i}$ and ${\bf e}_{j}$ are unit vectors characterizing
the orientations of the local
magnetic moments at sites $i$ and $j$, and the $J_{ij}$ denote the
effective exchange interactions between a pair of magnetic atoms,
provides a suitable tool to describe magnetic properties at finite
temperatures.
We can identify $\Delta$ with the energy resulting from a classical
Heisenberg model since the  corresponding energies in the FM and PM 
states are given by $E_{\rm FM}=-x^{2}\sum_{i\ne 0} J_{0i}^{\rm Mn,Mn}$ 
and $E_{\rm PM}=0$, respectively.
$J_{0i}^{\rm Mn,Mn}$ denote the exchange interactions between Mn-atoms 
and all other interactions are neglected.
On the other hand, the mean-field theory of the classical Heisenberg model
yields $k_{\rm B} {\tilde T}_{c}= 2x \sum_{i\ne 0} J_{0i}^{\rm Mn,Mn}/3$,
so that we have the following simple estimate,
\begin{equation}
k_{{\rm B}} {\tilde T}_{c}=2\Delta /3x \, .
\label{eq_TcDLM}
\end{equation}
It should be noted that non-magnetic As-antisites influence the Curie
temperature only indirectly via $\Delta$, which depends on the
concentrations $x$ and $y$ of Mn-atoms and As-antisites, respectively.

\subsection{ Effective on-site exchange parameters}
\label{EEP}

A more systematic approach to the evaluation of $T_{c}$ was developed 
by Liechtenstein and coworkers \cite{lie} and it is based on the
evaluation of magnetic excitation energies from the ground state
which are mapped onto the effective Heisenberg Hamiltonian. 
The Curie temperature is then obtained from its consequent statistical 
study.
A simpler approach, which avoids the direct evaluation of exchange
interactions, consists in the determination of the effective exchange 
field acting on the magnetic moment at this site.
 From this field the Curie temperature is estimated using the generalized 
molecular field theory \cite{lie,curieb}.
In the framework of the mean-field approximation adopted here are both 
approaches equivalent \cite{lie}, but the latter one is much simpler when 
applied to complex systems with few sublattices. 
Recently, we have applied this approach to evaluate Curie temperatures
of bulk ferromagnets \cite{curieb} as well as low-dimensional systems such 
as ultrathin films \cite{curies} and achieved a good agreement with 
available experimental data.

The basic quantity needed for a determination of $T_{c}$ is the
effective on-site exchange parameter for Mn-atoms, ${\bar J}_{i}^{\rm Mn}$, 
which represents the Weiss field acting on site $i$.
In the framework of the TB-LMTO approach we obtain \cite{lie,curieb}
\begin{equation}
{\bar J}_{i}^{\rm Mn} = - \frac{1}{4\pi} \, {\rm Im} \int_{C} \,
 {\rm tr}_L \, \Big[ \delta^{\rm Mn}_{i}(z) \, 
( {\bar g^{\rm Mn, \uparrow}}_{ii}(z) - 
{\bar g^{\rm Mn, \downarrow}}_{ii}(z) )
+ \delta^{\rm Mn}_{i}(z) \, {\bar g^{\rm Mn, \uparrow}}_{ii}(z) \,
 \delta^{\rm Mn}_{i}(z) \, {\bar g^{Mn, \downarrow}}_{ii}(z) \Big]
\, {\rm d} z \, .
\label{eq_Ji}
\end{equation}
Here ${\rm tr}_{L}$ denotes the trace over angular momenta $L=(\ell m)$,
the energy integration is performed in the upper half of the complex energy
plane over a contour $C$ starting below the bottom of the valence band
and ending at the Fermi energy, and $\delta^{\rm Mn}_{i}(z)=
P_{i}^{{\rm Mn},\uparrow}(z)-P_{i}^{{\rm Mn},\downarrow}(z)$, where
$P_{i}^{{\rm Mn},\sigma }(z)$ are the $L$-diagonal potential functions
of Mn atoms corresponding to the spin-index $\sigma$ 
($\sigma=\uparrow ,\downarrow $).
The quantity $\delta^{{\rm Mn}}_{i}(z)$ is proportional to the 
corresponding exchange splitting.
Finally, the quantities ${\bar{g}^{Mn, \sigma }}_{ii}(z)$ refer to 
site-diagonal blocks of the conditionally averaged Green function, 
namely, the average of the Green function over all configurations 
with a Mn-atom fixed at the site $i$ \cite{book}.
The quantity ${\bar J_i}^{\rm Mn}$ reflects the details of the electronic 
and lattice structures of the system, its carrier concentration, as well 
as the effect of various kinds of impurities on the resulting electronic 
structure.
The mean-field value of the Curie temperature is given by
\begin{equation}
k_{{\rm B}}T_{c}=\frac{2}{3} {\bar J_i}^{\rm Mn} \, .
\label{eq_Tc}
\end{equation}
The effective on-site exchange parameters corresponding to non-magnetic 
atoms are at least two-orders of magnitude smaller than 
${\bar J_i}^{\rm Mn}$. 
In Table~I we compare ${\bar T}_{c}$ calculated from expressions
(\ref{eq_TcDLM}) and (\ref{eq_Tc}): both approaches give similar
values.

\begin{figure}[tbp]
\epsfxsize=18cm \centerline{\epsffile{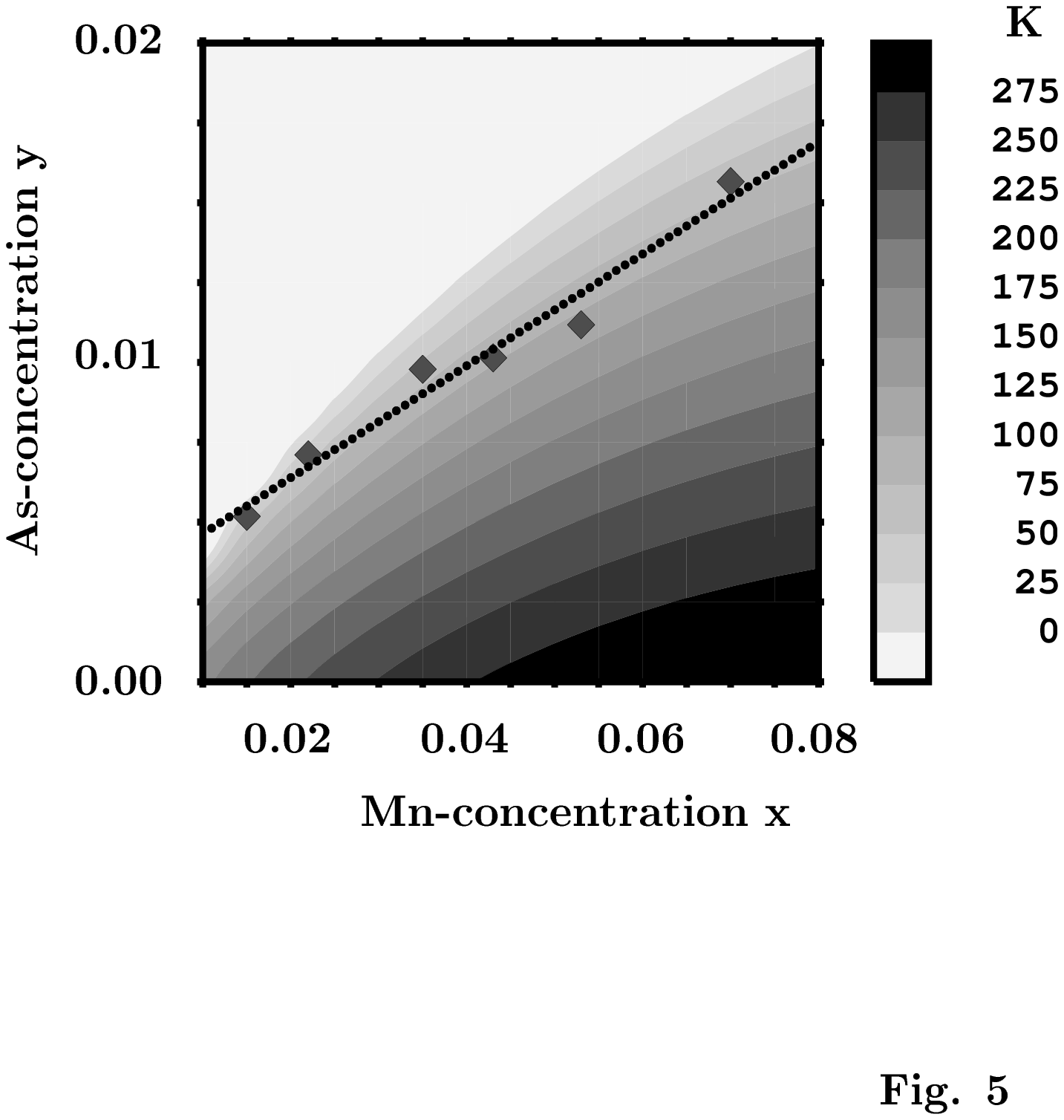}}
\vspace*{-7cm}
\caption{ Contour plot of the Curie temperature as a function of the
composition $(x,y)$ assuming the ferromagnetic ground state. The symbols
refer to experimental values [1] (see the text), the dotted line
represents a least-square fit to these data. }
\label{Fig.5}
\end{figure}

\begin{figure}[tbp]
\epsfxsize=11cm \centerline{\epsffile{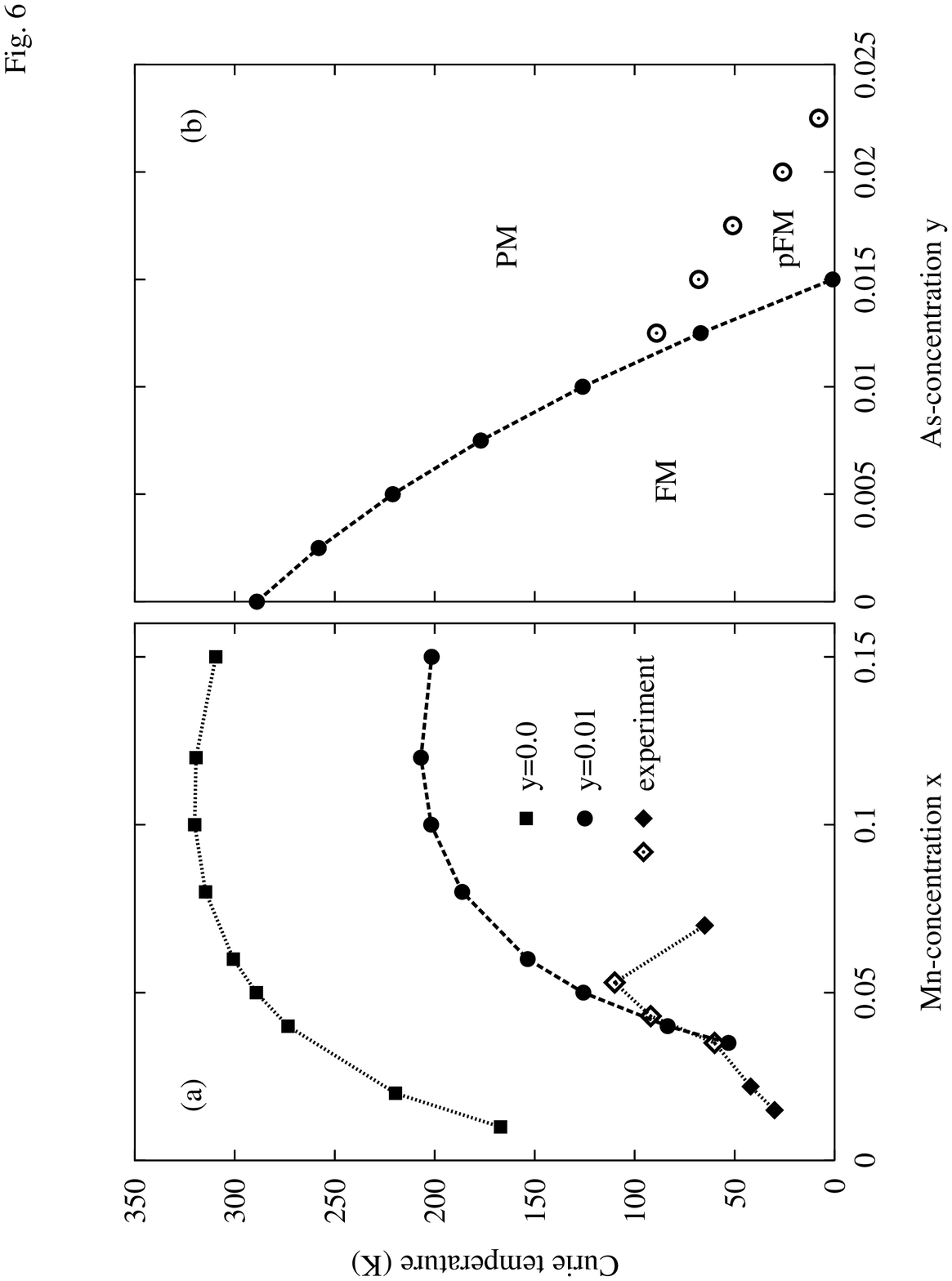}}
\vspace*{1cm}
\caption{ (a) Curie temperatures of (Ga$_{1-x-y}$Mn$_{x}$As$_{y}$)As in
the ferromagnetic state, Eq.~(\ref{eq_Tc}), as a function of $x$ for
$y=0$ and $y=0.01$ as compared to the experiment, Ref.~[1] (empty diamonds:
metallic ferromagnet, full diamonds: non-metallic samples);
(b) Curie temperatures of (Ga$_{0.95-y}$Mn$_{0.05}$As$_{y}$)As as a
function of $y$ in the ferromagnetic state, Eq.~(\ref{eq_Tc}) (full
circles) and in the partial ferromagnetic state (empty circles) which
is the ground state for $y > 0.011$. The symbols FM, pFM, and PM
label regions of parameters for which the ferromagnetic, partial
ferromagnetic, and paramagnetic states exist, respectively. }
\label{Fig.6}
\end{figure}

A general trend of Curie temperatures in the FM phase as a function of
the chemical composition $(x,y)$ is illustrated in Fig.~\ref{Fig.5}.
$T_{c}$ increases with $x$ in systems without As-antisites and reaches 
the room temperature at approximately $x=0.05$ and then decreases.
The same trend and similar values of $T_{c}$ were obtained recently
in the framework of the frozen-magnon approach applied to large ordered 
supercells \cite{sb}.
The Curie temperature for a fixed Mn-concentration is strongly reduced 
with increasing concentration of As-antisites.
This result clearly indicates a strong correlation between $T_{c}$ and 
the number of carriers known also from model studies \cite{kem,tckem}.
Detailed results for (Ga$_{1-x-y}$Mn$_{x}$As$_{y}$)As alloys without 
($y=0.0$) and with ($y=0.01$) As-antisites are presented in 
Fig.~\ref{Fig.6}a.
In both cases $T_{c}$ saturates for higher Mn-concentrations which in 
turn indicates that the effective exchange interactions 
${\bar{J}}_{0i}^{\rm Mn,Mn}$ should decrease with increasing Mn doping.
For $y=0.01$ the Curie temperature is reduced by approximately 
100~K$-$150~K over the whole range of Mn-concentrations as compared
to the case of $y=0.0$; the results are in a reasonably good agreement 
with experimental data for ferromagnetic metallic samples 
($0.035 < x < 0.055$) \cite{exp}. 

The concentration of As-antisites $y$ in Mn-enriched samples
is not well known from the experiment \cite{exp}.
The detailed knowledge of $T_{c}$ as a function of $x$ and $y$, 
Fig.~\ref{Fig.5}, allows to make an estimate of the relation
between $x$ and $y$ in such systems.
We have inserted the experimental points into Fig.~\ref{Fig.5}:
for each experimental $T_{c}$ corresponding to a given $x$ we have determined 
the concentration $y$ such that the calculated and experimental \cite{exp} 
Curie temperatures coincide.
Points obtained in this way follow approximately a straight line.
This allows to make an important conclusion, namely that the number of 
As-antisites increases with the concentration of Mn-atoms.
A recent evaluation of the formation energy of the As-antisite defect in 
GaMnAs \cite{jm} confirms this conclusion.

The dependence of $T_{c}$ on the concentration of As-antisites for 
a fixed Mn-concentration is presented in Fig.~\ref{Fig.6}b.
Full circles correspond to the FM state.
There is a monotonic decrease of $T_{c}$ with increasing $y$ due 
to the decreasing number of holes in the valence band which
mediate the ferromagnetic coupling between Mn atoms.
This result is in a qualitative agreement with predictions of the
KEM model \cite{kem,dietl,tckem}.
In the present case, however, the concentration of holes is not 
a free parameter but it is given by the chemical composition ($x,y$).
We have also determined $T_{c}$ for the pFM state which is the ground 
state for $y > 0.011$ (see Fig.~\ref{Fig.1}b).
We have now two magnetic atoms Mn$^{+}$ and Mn$^{-}$, but the Curie 
temperature $T_{c}$ can be still estimated from Eq.~(\ref{eq_Tc}).
It should be noted that now the larger of the two values 
${\bar J}_{i}^{\rm Mn^{\pm}}$ has to be inserted in (\ref{eq_Tc}) since 
the magnetism of such alloy is dominated by this constituent 
\cite{sakuma} (empty circles).
The transition between the FM and pFM states is manifested by the
change of the slope of the curve (see also Fig.~\ref{Fig.2}).
The regions of parameters for which the ferromagnetic, partial 
ferromagnetic, and paramagnetic phases exist are also shown.
It should be noted that in the pFM case the Curie temperature is shifted 
to higher values of $y$ than in the FM phase.

\subsection{ Effect of disorder on Curie temperature}
\label{DIS}

Further we investigate the effect of disorder, neglected in the framework 
of the KEM, on the Curie temperature.
For (Ga$_{0.95-y}$Mn$_{0.05}$As$_{y}$)As alloys with constant
Mn content is the effect of impurities in the framework of the KEM reduced 
simply to the change of the number of carriers.
The present theory includes the effect of disorder on the electronic 
structure and hence also on $T_c$ in the framework of the CPA. 
The Curie temperature is influenced by disorder via quantities 
${\bar g^{\rm Mn, \sigma}}_{ii}(z)$ in Eq.~(\ref{eq_Ji}) which depend
on the varying concentration of Mn-atoms and As-antisites.
We will demonstrate that the effect of disorder on $T_c$ is non-negligible.

To this end we have evaluated the Curie temperature as a function of the 
number of holes in the valence band while keeping the same reference density
of states, i.e., by employing the rigid-band model.
All quantities in (\ref{eq_Ji}) correspond to the reference system but
the Fermi energy is an adjustable parameter corresponding to a given
alloy composition.
The results are summarized in Table~\ref{Tab.2} assuming 
(Ga$_{0.95}$Mn$_{0.05}$)As as a reference system and are compared with 
selfconsistent calculations in which the effect of impurities is fully 
taken into account.
We will first consider the effect of As-antisites assuming 
(Ga$_{0.95-y}$Mn$_{0.05}$As$_{y}$)As alloys for which the number of holes in 
the valence band varies as $n_{h}=0.05 - 2y$ while the concentration of Mn 
atoms is kept constant.
The selfconsistent calculations and the rigid-band values follow the same 
trend but the latter ones are systematically smaller and quantitative 
differences increase with $y$: for $y=0.005$ and $y=0.01$ are the rigid-band
values 88\% and 62\% of the selfconsistent ones, respectively.

It is also possible to use the same approach for estimating $T_c$ for
(Ga$_{1-x}$Mn$_{x}$)As alloys without As-antisites.
In this case, however, we have to scale the calculated rigid-band values
in the spirit of the KEM model by a factor $x_{\rm ref}/x$, where the
reference Mn-concentration $x_{\rm ref}=0.05$.
The results are similar to the previous case: we observe the same trends
in both cases but the rigid-band values are lower than in the case in which 
the effect of disorder is fully taken into the account.
We can conclude that the neglect of disorder does not change the qualitative 
trends of the calculated $T_c$ as a function of Mn- and As-concentrations.
On the other hand, the disorder leads in general to an enhancement of $T_c$
as compared to the model in which the effect of disorder is treated
approximately in the framework of the rigid-band model.
This is not surprising as the rigid-band model is not able to describe
properly the effect of As- and Mn-impurities which both represent a strong
perturbation of the electronic structure.

\section{CONCLUSIONS}
\label{sec_Concl}

Based on first-principles calculations we have investigated the effect 
of As-antisites on the electronic and magnetic properties of 
(Ga,Mn)As alloys.
In addition to chemical disorder magnetic disorder was described 
using the disordered local moment model.
Both the chemical and the magnetic disorder are treated within the
framework of the coherent potential approximation.
The mean-field Curie temperatures were estimated from the effective 
exchange fields acting on a given magnetic atom as well as from total 
energy differences between the ferromagnetic and paramagnetic state and 
showed a reasonable agreement between both approaches.

The main results can be summarized as follows: 
(i) there exists an additional phase with partially ordered 
local magnetic moments which separates the FM and PM states in the 
magnetic phase diagram. This phase is stabilized by As-antisites and 
has a reduced magnetization;
(ii) our detailed study of the local magnetic moments justifies
{\it a posteriori} the validity of the disordered local moment model 
for determination of magnetic properties of DMS;
(iii) a strong reduction of the Curie temperature occurs with increasing
concentration of the As-antisites;
(iv) with increasing concentration of Mn-atoms the calculations indicate 
a reduction of the exchange interactions between Mn atoms;
(v) a comparison of the calculated and the measured concentration 
dependences of the Curie temperature indicates a correlation between 
the concentrations of Mn-impurities and As-antisites, namely an increase 
of the donor concentration with an increase of the Mn-content; and
{\bf 
(vi) a proper account of the influence of impurities on the electronic 
structure is needed for a quantitative estimate of the Curie temperature, 
in particular, the Curie temperature is reduced if the effect of disorder 
on the electronic structure is neglected.}

\section*{Acknowledgements}

The financial support provided by the Grant Agency of the Czech Republic
(No.\ 202/00/0122), the Grant Agency of the Academy of Sciences of the
Czech Republic (No.\ A1010203, A1010214), the Czech Ministry of Education,
Youth, and Sports (COST P5.30, MSM 113200002), the CMS Vienna (GZ 45.504),
and the RTN {\it Computational Magnetoelectronics} (HPRN-CT-2000-00143) is
gratefully acknowledged.

\newpage

\begin{table}[tbp]
\caption{ Curie temperatures [K] for the ferromagnetic state. The values
${\tilde T}_c$ and $T_c$ are determined from Eqs.~(\ref{eq_TcDLM}) and
(\ref{eq_Tc}), respectively.
Note that the ground state for $y > 0.01$ is a partial ferromagnet. }
\label{Tab.1}
\begin{center}
\begin{tabular}{dccdcc}
 \multicolumn{3}{c}{(Ga$_{1-x}$Mn$_x$)As} &
 \multicolumn{3}{c}{(Ga$_{0.95-y}$Mn$_{0.05}$As$_y$)As} \\
  $x$  & ${\tilde T}_c$ & $T_c$ & $y$ & ${\tilde T}_c$ & $T_c$  \\ \hline
  0.02 &  281.8 K  &  219.6 K & 0.0    & 346.8 K & 289.2 K  \\
  0.04 &  332.1 K  &  273.4 K & 0.0025 & 328.0 K & 258.1 K  \\
  0.06 &  358.2 K  &  300.7 K & 0.005  & 282.3 K & 220.6 K  \\
  0.08 &  373.5 K  &  314.6 K & 0.0075 & 239.5 K & 176.6 K  \\
  0.10 &  381.9 K  &  320.1 K & 0.01   & 188.2 K & 125.7 K  \\
  0.12 &  385.3 K  &  319.4 K & 0.0125 &   $-$   &  $-$    \\
  0.15 &  383.7 K  &  309.5 K & 0.015  &   $-$   &  $-$    \\
 \end{tabular}
\end{center}
\end{table}

\begin{table}[tbp]
\caption{ Curie temperatures [K] for the ferromagnetic state. The values
$T_c$ are determined from (\ref{eq_Tc}) while $T_{c}(n_{h})$ is its 
rigid-band counterpart (see Sec.~\ref{DIS}) determined for the reference
Mn-concentration $x_{\rm ref}=0.05$. Both values coincide for the reference
case of (Ga$_{0.95}$Mn$_{0.05}$)As.}
\label{Tab.2}
\begin{center}
\begin{tabular}{dccdcc}
 \multicolumn{3}{c}{(Ga$_{1-x}$Mn$_x$)As} &
 \multicolumn{3}{c}{(Ga$_{0.95-y}$Mn$_{0.05}$As$_y$)As} \\
  $x$  & $(x_{\rm ref}/x)\; T_{c}(n_{h})$ & $T_c$ & $y$ & $T_{c}(n_{h})$ & $T_c$  \\ \hline
  0.03 &  130 K  &  251 K & 0.0    & 289 K & 289 K  \\
  0.04 &  244 K  &  273 K & 0.005  & 195 K & 221 K  \\
  0.05 &  289 K  &  289 K & 0.0075 & 142 K & 177 K  \\
  0.06 &  285 K  &  301 K & 0.01   &  79 K & 126 K  \\
 \end{tabular}
\end{center}
\end{table}

\newpage
\vspace*{4cm}

\newpage
\vspace*{4cm}

\newpage
\vspace*{5cm}

\newpage
\vspace*{-2cm}

\newpage
\vspace*{5cm}

\end{document}